\documentclass[prd,twocolumn,footinbib,showpacs]{revtex4}

\begin{document}

\title{Interacting dark energy in $f(R)$ gravity} 
\author{Nikodem J. Pop\l awski}
\affiliation{Department of Physics, Indiana University,
727 East Third Street, Bloomington, Indiana 47405, USA}
\date{\today}

\begin{abstract}

The field equations in $f(R)$ gravity derived from the Palatini variational
principle and formulated in the Einstein conformal frame yield a cosmological
term which varies with time.
Moreover, they break the conservation of the energy--momentum tensor for
matter, generating the interaction between matter and dark energy.
Unlike phenomenological models of interacting dark energy, $f(R)$ gravity
derives such an interaction from a covariant Lagrangian which is a function of
a relativistically invariant quantity (the curvature scalar $R$).
We derive the expressions for the quantities describing this interaction
in terms of an arbitrary function $f(R)$, and
examine how the simplest phenomenological models of a variable
cosmological constant are related to $f(R)$ gravity.
Particularly, we show that $\Lambda c^2=H^2(1-2q)$ for
a flat, homogeneous and isotropic, pressureless universe.
For the Lagrangian of form $R-1/R$, which is the simplest way of introducing
current cosmic acceleration in $f(R)$ gravity, the predicted matter--dark energy
interaction rate changes significantly in time, and its current value is
relatively weak (on the order of $1\%$ of $H_0$),
in agreement with astronomical observations.

\end{abstract}

\maketitle

\section{Introduction}

Einstein's general relativity is based on the Lagrangian that is
a linear function of the Riemann curvature scalar.
The linearity of the gravitational action with respect to
the curvature results in the Einstein equations of the
gravitational field that are second-order differential equations for
the metric tensor~\cite{LL}.
The Einstein equations applied to a homogeneous and isotropic spacetime
lead to the Friedmann equations describing an expanding universe with
a positive deceleration.
The most accepted explanation of the observed cosmic
acceleration~\cite{univ} is that the universe is dominated by dark
energy (quintessence)~\cite{DE1,DE2}.
The simplest way of introducing dark energy into general relativity is
to add a cosmological constant $\Lambda$ to the curvature scalar $R$ in
the Lagrangian (the $\Lambda CDM$ model):
\begin{equation}
S_g=-\frac{1}{2\kappa c}\int d^4 x\sqrt{-g}[R+2\Lambda].
\label{cosm}
\end{equation}

However, it is also possible to modify the curvature dependence of the
gravitational action to obtain the field equations that allow accelerated
expansion. 
A particular class of alternative theories of gravity that has recently 
attracted a lot of interest is that of the $f(R)$ gravity 
models, in which the gravitational Lagrangian is a nonlinear function of
$R$~\cite{fR}. 
It has been shown that current cosmic acceleration may originate from the
addition of a term $R^{-1}$ (or other negative powers of $R$) to the
Einstein--Hilbert Lagrangian $R$~\cite{acc}.
As in general relativity, $f(R)$ gravity theories obtain the field equations
by varying the total action for both the field and matter. 

In the literature, there are two approaches on how to perform the variation.
We use the metric--affine (Palatini) variational principle,
according to which the metric and connection are considered as 
geometrically independent quantities, and the action is varied 
with respect to both of them. 
The other one is the metric (Einstein--Hilbert) variational principle, 
according to which the action is varied with respect to the metric 
tensor $g_{\mu\nu}$, and the affine connection coefficients are
the Christoffel symbols of $g_{\mu\nu}$.
The field equations in the Palatini formalism are second-order 
differential equations, while for metric theories they are 
fourth-order.
Both approaches give the same result 
only if we use the standard Einstein--Hilbert action~\cite{Schr}.

An interaction between ordinary matter and dark energy has been introduced in
the form of a time dependent cosmological constant $\Lambda(t)$~\cite{var}.
A decreasing (decaying) $\Lambda$ gives a positive rate of the entropy production
and could explain the observed large entropy of the universe.
The cosmological constant can depend on the cosmic time directly~\cite{t} or
through other cosmological variables such as the scale factor $a$ or the Hubble parameter $H$.
Dimensional arguments lead to $\Lambda\propto a^{-2}$~\cite{R2a,R2b},
$\Lambda\propto H^2$~\cite{H}, or both~\cite{HR}.
A more general form $\Lambda\propto a^{-m}$, where $m=\mbox{const}$~\cite{Rm},
is predicted by holographic cosmology~\cite{hol1,hs,hol2}.
The cosmological term can also involve other
quantities such as the deceleration parameter or temperature.
Ref.~\cite{OC} lists and reviews phenomenological models with a variable cosmological term,
and examines evolution of the scale factor in these models.

The field equations in $f(R)$
gravity formulated in the Einstein conformal frame not only yield
an effective cosmological term, but also break the
conservation of the energy--momentum tensor for matter~\cite{Niko1},
generating the interaction between matter and dark energy.
Therefore, $f(R)$ gravity provides a physical explanation for the
exchange of energy between cosmological term and other forms of matter.
Unlike phenomenological models of interacting dark energy, and
like Brans--Dicke cosmologies with the cosmological term depending
on the scalar field~\cite{Ber}, it derives such an interaction from
a covariant Lagrangian in which the cosmological constant is a function
of a relativistically invariant quantity ($R$).
Furthermore, $f(R)$ gravity in the Einstein frame generates a variation
of the gravitational constant with the cosmological time, as in the
large number hypothesis of Dirac~\cite{Dirac}.

The aim of this paper is to find how this interaction depends on the function $f(R)$.
In Sec.~II we derive the equations of field in $f(R)$ gravity using
the Palatini variational principle.
In Sec.~III we apply these equations to a homogeneous and isotropic,
flat universe filled with dust, and obtain the expressions describing
the interaction between matter and dark energy.
We also examine how the simplest phenomenological models of a variable
cosmological constant are related to $f(R)$ gravity.
In Sec.~IV we give numerical predictions for the case $f(R)=R+\mbox{const.}\times R^{-1}$, 
which is the simplest way of introducing current cosmic acceleration in
$f(R)$ gravity~\cite{acc},
and compare them with other models of interacting dark energy.
The results are summarized in Sec.~V.

\section{The field equations}

The action for an $f(R)$ gravity is given by~\cite{Niko1,eq2}
\begin{equation}
S_J=-\frac{1}{2\kappa c}\int d^4 
x\bigl[\sqrt{-\tilde{g}}f(\tilde{R})\bigr] + 
S_m(\tilde{g}_{\mu\nu},\psi).
\label{action1}
\end{equation}
Here, $\sqrt{-\tilde{g}}f(\tilde{R})$ is a Lagrangian density that depends 
on the curvature scalar 
$\tilde{R}=R_{\mu\nu}(\Gamma^{\,\,\lambda}_{\rho\,\sigma})\tilde{g}^{\mu\nu}$, 
$S_m$ is the action for matter represented 
symbolically by $\psi$ and assumed to be independent of the connection
$\Gamma^{\,\,\lambda}_{\rho\,\sigma})$, and the connection is assumed
to be symmetric (no torsion).
The variables describing an $f(R)$ Lagrangian are said to form
the Jordan conformal frame ({\it JCF})~\cite{JE}.
Tildes indicate quantities calculated in this frame, e.g., 
$\tilde{g}_{\mu\nu}$ is the JCF metric tensor.

Variation of the action $S_J$ with respect to $\tilde{g}_{\mu\nu}$ 
yields the field equations
\begin{equation}
f'(\tilde{R})R_{\mu\nu}-\frac{1}{2}f(\tilde{R})\tilde{g}_{\mu\nu}=\kappa 
T_{\mu\nu},
\label{field1}
\end{equation} 
where the dynamical energy--momentum tensor of matter $T_{\mu\nu}$ is
generated by the JCF metric tensor:
\begin{equation}
\delta S_m=\frac{1}{2c}\int d^4 x\sqrt{-\tilde{g}}\,T_{\mu\nu}\delta\tilde{g}^{\mu\nu}.
\label{EMT}
\end{equation} 
The prime denotes the derivative of the function $f(\tilde{R})$ with
respect to $\tilde{R}$.
From variation of $S_J$ with 
respect to the connection $\Gamma^{\,\,\rho}_{\mu\,\nu}$
it follows that this connection is given by the Christoffel 
symbols of the conformally transformed metric~\cite{fR} 
\begin{equation}
g_{\mu\nu}=f'(\tilde{R})\tilde{g}_{\mu\nu}.
\label{conf}
\end{equation}
The metric $g_{\mu\nu}$ defines the Einstein conformal frame ({\it ECF}),
in which the connection is metric compatible~\cite{JE}.

A transition from the JCF (modified gravity) to the ECF (general relativity)
in $f(R)$ gravity is possible if $S_m$ does not contain torsion~\cite{eq2}.
Such a transition is also possible for more general theories in which the
gravitational Lagrangian depends on the Ricci tensor, but not on the Weyl
tensor~\cite{eq3}.
In the Jordan frame, the connection is metric incompatible unless $f(R)=R$.
We regard the ECF metric tensor as physical, although whether
it is true or not should be ultimately decided by experiment or observation.
Since the physical equivalence between both frames is an open
problem as well~\cite{eqv}, so is the answer to the question whether and how
the results of this paper would change if the JCF was physical.

In the Einstein frame, the action~(\ref{action1}) becomes the standard 
general-relativistic action of the gravitational field interacting
with an additional nondynamical scalar field~\cite{eq1,Niko1}:
\begin{eqnarray}
& & S_E=-\frac{1}{2\kappa c}\int d^4 x\sqrt{-g}\bigl[R-2V(\tilde{R})\bigr] \nonumber \\
& & +S_m([f'(\tilde{R})]^{-1}g_{\mu\nu},\psi),
\label{action2}
\end{eqnarray}
where $R=R_{\mu\nu}(\Gamma^{\,\,\lambda}_{\rho\,\sigma})g^{\mu\nu}$ is
the Riemannian curvature scalar of the metric $g_{\mu\nu}$, and
$V(\tilde{R})$ is the effective potential,
\begin{equation}
V(\tilde{R})=\frac{\tilde{R} f'(\tilde{R})-f(\tilde{R})}{2[f'(\tilde{R})]^2}.
\label{pot}
\end{equation}
The curvature scalars in both frames are related by
\begin{equation}
R=f'(\tilde{R})\tilde{R},
\label{resc}
\end{equation}
which follows from~(\ref{conf}).

Variation of the action~(\ref{action2}) with respect to 
$g_{\mu\nu}$ gives the equation of the gravitational field in
the Einstein frame~\cite{field,Niko1}: 
\begin{equation}
R_{\mu\nu}-\frac{1}{2}Rg_{\mu\nu}=\frac{\kappa 
T_{\mu\nu}}{f'(\tilde{R})}-V(\tilde{R})g_{\mu\nu}.
\label{EOF1}
\end{equation}
Eqs.~(\ref{resc}) and~(\ref{EOF1}) yield an algebraic relation
\begin{equation}
\tilde{R} f'(\tilde{R})-2f(\tilde{R})=\kappa Tf'(\tilde{R}),
\label{struc}
\end{equation}
from which we obtain $\tilde{R}$ as a function of $T=T_{\mu\nu}g^{\mu\nu}$
for a given $f(\tilde{R})$~\footnote{
Except for the ``singular'' case $f(\tilde{R})=\tilde{R}^2$, for which $T$ vanishes identically and $\tilde{R}$ is undetermined.}.
If $T=0$ (vacuum or radiation) then $\tilde{R}=\mbox{const.}$ and
the solution of the field equation
is an empty spacetime with a cosmological constant~\cite{Schr}.
Substituting $\tilde{R}$ into~(\ref{EOF1}) leads
to a final relation between the geometrical tensors and the energy--momentum
tensor,
\begin{equation}
R_{\mu\nu}-\frac{1}{2}Rg_{\mu\nu}=\kappa_r(T)T_{\mu\nu}+\Lambda(T)g_{\mu\nu},
\label{EOF2}
\end{equation}
with a running gravitational coupling
$\kappa_r(T)=\kappa[f'(\tilde{R}(T))]^{-1}$
and a variable cosmological term $\Lambda(T)=-V(\tilde{R}(T))$:
\begin{equation}
\Lambda(\tilde{R})=\frac{f(\tilde{R})-\tilde{R} f'(\tilde{R})}{2[f'(\tilde{R})]^2}.
\label{cc}
\end{equation}
Such a relation is, in general, nonlinear and depends on the form of the
function $f(\tilde{R})$. 
Moreover, the time dependence of $T$ makes the quantities $\kappa_r$
and $\Lambda$ variable, although they are not associated with any dynamical
quintessence scalar field as in~\cite{DE1,dyn}.
Concluding, Palatini $f(R)$ gravity does not change the structure of
Einstein's general relativity but introduces a varying gravitational constant~\cite{Dirac},
as well as a term acting like a cosmological constant which varies
with time.
It modifies the gravitational field inside material objects, while
the metric in vacuum (such as the Schwarzschild solution for spherically
symmetric systems) is the same as that in general relativity.

\section{The interaction between matter and dark energy}

The Bianchi identity applied to~(\ref{EOF1}) gives
\begin{equation}
T_{\mu\nu}^{\,\,\,\,\,;\nu}=\tilde{R}^{,\nu}f''(\tilde{R})\Bigl(\frac{T_{\mu\nu}}{f'(\tilde{R})}+\frac{[2f(\tilde{R})-\tilde{R} f'(\tilde{R})]g_{\mu\nu}}{2\kappa[f'(\tilde{R})]^2}\Bigr).
\label{conserv}
\end{equation} 
This relation means that the energy--momentum tensor in the Einstein frame is
{\em not} covariantly conserved, unless $f(R)=R$ or $T=0$~\cite{Niko2}.
We can write the field equation~(\ref{EOF1}) as
\begin{equation}
R_{\mu\nu}-\frac{1}{2}Rg_{\mu\nu}=\kappa(T_{\mu\nu}^m+T_{\mu\nu}^{\Lambda}),
\label{EOF3}
\end{equation}
where $T_{\mu\nu}^m=T_{\mu\nu}$.
This defines the dark energy--momentum tensor,
\begin{equation}
T_{\mu\nu}^{\Lambda}=\frac{\Lambda(\tilde{R})}{\kappa}g_{\mu\nu}+\frac{1-f'(\tilde{R})}{f'(\tilde{R})}T_{\mu\nu}.
\label{darkt}
\end{equation}
From Eq.~(\ref{EOF3}) it follows that matter and dark energy
form together a system that has a conserved four-momentum.
Therefore, we can speak of an {\em interaction} between matter and dark
energy~\cite{dec}.
In the Jordan frame of $f(R)$ gravity, the energy--momentum tensor is always
conserved~\cite{Koi} as a direct consequence of Eq.~(\ref{EMT}) for the
case of an infinitesimal translation~\cite{LL}.

For a flat~\cite{Gold} Robertson--Walker universe with pressureless matter
(dust) $p_m=0$ and in the comoving frame of reference, Eq.~(\ref{conserv})
gives the first Friedmann equation,
\begin{equation}
f'(\tilde{R})\frac{d}{dt}\biggl[\frac{\epsilon_m(\tilde{R})}{f'(\tilde{R})}-\frac{V(\tilde{R})}{\kappa}\biggr]+3H(\tilde{R})\epsilon_m(\tilde{R})=0,
\label{cons1}
\end{equation}
where the Hubble parameter is given by~\cite{Niko1}
\begin{equation}
H(\tilde{R})=\frac{c}{f'(\tilde{R})}\sqrt{\frac{\tilde{R} f'(\tilde{R})-3f(\tilde{R})}{6}}.
\label{Hub}
\end{equation}
The matter energy density $\epsilon_m(\tilde{R})=T(\tilde{R})$ is obtained from~(\ref{struc}),
\begin{equation}
\epsilon_m=\frac{\tilde{R} f'(\tilde{R})-2f(\tilde{R})}{\kappa f'(\tilde{R})}.
\label{energyden}
\end{equation}
Eqs.~(\ref{cons1}) and~(\ref{Hub}) give the time evolution of $\tilde{R}$,
\begin{equation}
\dot{\tilde{R}}=\frac{\sqrt{6}c(\tilde{R} f'-2f)\sqrt{\tilde{R} f'-3f}}{2f'^2+\tilde{R} f'f''-6ff''},
\label{phidot}
\end{equation}
and we wrote $f=f(\tilde{R})$ for brevity.
The equation of continuity~(\ref{cons1}) can be written as
\begin{equation}
\dot{\epsilon}_m+3H\epsilon_m=Q,
\label{cons2}
\end{equation}
where
\begin{equation}
Q=\frac{\dot{\tilde{R}}f''(\tilde{R})[\tilde{R} f'(\tilde{R})-2f(\tilde{R})]}{2\kappa[f'(\tilde{R})]^2}.
\label{Q}
\end{equation}
The quantity $Q$ describes the interaction between matter and dark
energy~\cite{int1}, and vanishes for the general-relativistic case $f(R)=R$.

In the comoving frame of reference, Eq.~(\ref{darkt}) reads
\begin{eqnarray}
& & \epsilon_{\Lambda}=\frac{\Lambda}{\kappa}+\frac{1-f'(\tilde{R})}{f'(\tilde{R})}\epsilon_m,
\label{dark1} \\
& & p_\Lambda=-\frac{\Lambda}{\kappa}.
\label{dp}
\end{eqnarray}
The definitions in Eq.~(\ref{darkt}) were chosen so that $\Omega_m+\Omega_{\Lambda}=\epsilon_m/\epsilon_c+\epsilon_{\Lambda}/\epsilon_c=1$,
where $\epsilon_c=3H^2/(\kappa c^2)$ is the critical energy density,
\begin{equation}
\epsilon_c=\frac{\tilde{R}f'(\tilde{R})-3f(\tilde{R})}{2\kappa [f'(\tilde{R})]^2}.
\label{crit}
\end{equation}
We also find the equation of continuity for dark energy,
\begin{equation}
\dot{\epsilon}_{\Lambda}+3H(\epsilon_{\Lambda}+p_{\Lambda})=-Q,
\label{dark2}
\end{equation}
which has the same form as in~\cite{int1}.
Eqs.~(\ref{cons2}) and~(\ref{dark2}) are a realization of the fact that
the system matter--dark energy is closed.
The interaction term $Q$ corresponds to the production of
particles from quintessence.

In $f(R)$ gravity, the dark energy density is given by
\begin{equation}
\epsilon_{\Lambda}=\frac{\tilde{R}f'(\tilde{R})-3f(\tilde{R})-2f'(\tilde{R})[\tilde{R}f'(\tilde{R})-2f(\tilde{R})]}{2\kappa[f'(\tilde{R})]^2}.
\label{dark}
\end{equation}
Consequently, the rate of interaction
$\Gamma=Q/\epsilon_{\Lambda}$~\cite{int1} equals
\begin{equation}
\Gamma=\frac{\sqrt{6}cf''(\tilde{R}f'-2f)^2\sqrt{\tilde{R}f'-3f}}{[\tilde{R}f'-3f-2f'(\tilde{R}f'-2f)](2f'^2+\tilde{R}f'f''-6ff'')},
\label{gam}
\end{equation}
and the nondimensional ratio $\gamma=\Gamma/H$ takes the form
\begin{equation}
\gamma=\frac{6f'f''(\tilde{R}f'-2f)^2}{[\tilde{R}f'-3f-2f'(\tilde{R}f'-2f)](2f'^2+\tilde{R}f'f''-6ff'')}.
\label{rat}
\end{equation}
Finally, the ratio $w_\Lambda=p_\Lambda/\epsilon_\Lambda$ is given by
\begin{equation}
w_\Lambda=\frac{\tilde{R}f'-f}{\tilde{R}f'-3f-2f'(\tilde{R}f'-2f)}.
\label{w}
\end{equation}
The case $w_\Lambda=\mbox{const}$ was examined, for example, in~\cite{cw}.
In $f(R)$ gravity, both $\gamma$ and $w_\Lambda$ are {\it not}
constant, in general.

We now examine the consistence of the $f(R)$ description
of interacting dark energy with some of the phenomenological models studied in~\cite{OC}.
We begin with the decay law $\Lambda=\beta H^2/c^2$, where $\beta$ is a
nondimensional and positive (since $\Lambda$ decreases with time) constant. 
This condition follows from the dimensional relations $t_H\sim H^{-1}$ and
$\Lambda\sim l^{-2}_{Pl}(t_{Pl}/t_H)^2$, where $t_H$ is the Hubble time,
and $l_{Pl}$ and $t_{Pl}$ are respectively the Planck length and time.
It can also be deduced from Dirac's large number hypothesis~\cite{H},
and the holographic principle~\cite{pz}. 
Using Eqs.~(\ref{cc}) and~(\ref{Hub}) we obtain
\begin{equation}
\frac{\tilde{R}f'(\tilde{R})-f(\tilde{R})}{\tilde{R}f'(\tilde{R})-3f(\tilde{R})}=-\frac{\beta}{3},
\label{ex_H}
\end{equation}
which has the solution in the form of a power function, $f(\tilde{R})\sim \tilde{R}^{3(1+\beta)/(3+\beta)}$.
Therefore, the relation $\Lambda\propto H^2$ is inconsistent with $f(R)$ gravity 
models which assume $f(R)=R+\varepsilon g(R)$, where $g(R)$ is some
nonlinear function of $R$ and $\varepsilon$ is a small quantity.
Such a form of $f(R)$ is required for the deviations from general
relativity to be compatible with solar system experiments~\cite{small}.

Considering $\Lambda\propto H^2$ is equivalent~\cite{H}, at the level of the
Einstein equations in general-relativistic cosmology, to the constancy of the parameter~\cite{x}
\begin{equation}
x=\frac{\epsilon_{\Lambda}}{\epsilon_m+\epsilon_{\Lambda}},
\label{x1}
\end{equation}
i.e. the constancy of the ratio $r=\epsilon_m/\epsilon_\Lambda$~\cite{xx,hl}.
This is not the case in $f(R)$ gravity, where this parameter is given by
\begin{equation}
x=\frac{\tilde{R}f'-3f-2f'(\tilde{R}f'-2f)}{\tilde{R}f'-3f},
\label{x2}
\end{equation}
and its constancy leads to a first-order differential equation for $f(\tilde{R})$ which has no power-function solutions if $x\neq0$ and $x\neq1$~\footnote{
The case $x=0$ corresponds to the general-relativistic form $f(\tilde{R})=\tilde{R}$.
The case $x=1$ corresponds to the function $f(\tilde{R})=\tilde{R}^2$, for which
Eq.~(\ref{struc}) yields $T=0$ (pure radiation or empty universe).}.
A more general case $r\propto a^{-\xi}$, where $\xi=\mbox{const}$,
was examined in~\cite{Maj}.
The condition $\Lambda\propto H^2$ has also been used in~\cite{int2},
together with the constancy of $\gamma$ (and $r$).
The latter leads to a complicated, nonlinear second-order differential equation
for the function $f(\tilde{R})$.
A similar equation is obtained for the decay law $\gamma\propto r$
used in~\cite{cw,delt}.
We conclude that $f(R)$ gravity does not favor theories which constrain
$Q$, $\Gamma$, or $\gamma$.
Lastly, we note that a more general condition $\Lambda\propto H^n$, where
$n=\mbox{const}\neq2$, has no power-function solutions for $f(\tilde{R})$ as well. 

From the dependence of the redshift on $\tilde{R}$~\cite{Niko2} we obtain
\begin{equation}
a^{-3}=\frac{b[\tilde{R} f'(\tilde{R})-2f(\tilde{R})]}{[f'(\tilde{R})]^2},
\label{redsh}
\end{equation}
where the constant $b$ is given by
\begin{equation}
b=\frac{[f'(\tilde{R}_0)]^2}{a^3_0[\tilde{R}_0f'(\tilde{R}_0)-2f(\tilde{R}_0)]},
\label{b}
\end{equation}
and the subscript $0$ denotes the corresponding present value.
Therefore, the decay law $\Lambda\propto a^{-m}$, where $m=\mbox{const}$, 
yields
\begin{equation}
\frac{[f'(\tilde{R})]^{(\frac{2m}{3}-2)}[f(\tilde{R})-\tilde{R}f'(\tilde{R})]}{[\tilde{R}f'(\tilde{R})-2f(\tilde{R})]^{\frac{m}{3}}}=\delta=\mbox{const}.
\label{ex_Rm}
\end{equation}
If $m=3$ then Eq.~(\ref{ex_Rm}) has the power-function solution, $f(\tilde{R})\sim \tilde{R}^{(1-2\delta)/(1-\delta)}$.
The positivity of $\epsilon_\Lambda$ and $\Lambda$ requires $\delta<0$.
For $m=2$~\cite{R2a,R2b}, $m=4$~\cite{Rm,hol2,x}, or other values of $m$~\cite{OC}, Eq.~(\ref{ex_Rm}) is quite complicated.

We now express the condition $\Lambda\propto q$~\cite{OC}, where $q$ is the
deceleration parameter, in $f(R)$ gravity.
Using the formula for $q$~\cite{Niko2},
\begin{equation}
q(\tilde{R})=\frac{2\tilde{R}f'(\tilde{R})-3f(\tilde{R})}{\tilde{R}f'(\tilde{R})-3f(\tilde{R})},
\label{q}
\end{equation}
we find
\begin{equation}
\frac{(\tilde{R}f'-f)(\tilde{R}f'-3f)}{f'^2(2\tilde{R}f'-3f)}=\mbox{const}.
\label{ex_q}
\end{equation}
The only power function obeying this condition is $f(\tilde{R})=\tilde{R}^2$,
which is excluded by Eq.~(\ref{struc}).
Concluding, there is no simple correspondence between $f(R)$ gravity and
the phenomenological, power-law models examined in~\cite{OC} which relate
the cosmological constant to any (one) cosmological quantity from the set ($H$, $q$, $a$).
In order to find the function $f(\tilde{R})$ which corresponds to such an
expression for $\Lambda$, we need to solve a complicated, nonlinear differential equation.

However, using Eqs.~(\ref{Hub}),~(\ref{redsh}), and~(\ref{q}), we can compose
the exact cosmological term~(\ref{cc}) from any two cosmological paremeters
from the set ($H$, $q$, $a$), which gives three combinations.
We find the following relations:
\begin{eqnarray}
& & \Lambda=\frac{H^2}{c^2}(1-2q),
\label{genH} \\
& & \Lambda=\frac{3H^2}{c^2}-\frac{a^{-3}}{b},
\label{genRH} \\
& & \Lambda=\frac{1-2q}{2(1+q)}\,\frac{a^{-3}}{b}.
\label{Rq}
\end{eqnarray}
All three are satisfied in $f(R)$ gravity, but we only need one to compare
$f(R)$ gravity with a particular phenomenological model. 
The first equation generalizes the law $\Lambda\propto H^2$ in the
sense that the proportionality constant depends on the
deceleration parameter, $\beta=1-2q$.
Eq.~(\ref{genH}) has the same form for an arbitrary function $f(R)$,
and is an important prediction of the Palatini $f(R)$ gravity~\footnote{
Eqs.~(\ref{dp}) and~(\ref{genH}) yield the relation $p_\Lambda=-\frac{H^2}{\kappa c^2}(1-2q)$.
In general-relativistic cosmology without the cosmological constant,
we have $p_m=-\frac{H^2}{\kappa c^2}(1-2q)$.}.
The two other relations involve the constant $b$ which depends on the function
$f(\tilde{R})$ and the present value $\tilde{R}_0$, i.e. they are model
dependent.
Eq.~(\ref{genRH}) is similar to the form assumed in~\cite{HR}, except
we have $a^{-3}$ instead of $a^{-2}$.
We note that if $\Lambda$ depends on $q$ then the relation $q=1/2$,
which holds for the matter era, yields $\Lambda=0$.
 
In general-relativistic cosmology, some phenomenological laws are dynamically
equivalent to each other~\cite{wm,equi}.
For example, the relations $\Lambda\propto H^2$, $\Lambda\propto qH^2$~\cite{arb}, and $\Lambda\propto\epsilon_m$~\cite{vish} give the same cosmological
evolution~\cite{equi}.
Since the decay law $\Lambda\propto H^2$ appears to be inconsistent with current constraints
on the dark energy equation of state~\cite{hs,wm}, so do the other equivalent models. 
In $f(R)$ gravity, the condition $\Lambda=\mu qH^2/c^2$,
where $\mu=\mbox{const}$, becomes
\begin{equation}
\frac{\tilde{R}f'(\tilde{R})-f(\tilde{R})}{2\tilde{R}f'(\tilde{R})-3f(\tilde{R})}=-\frac{\mu}{3}.
\label{ex_Hq}
\end{equation}
Eq.~(\ref{ex_Hq}) has the solution in the form of a power function, $f(\tilde{R})\sim \tilde{R}^{3(1+\mu)/(3+2\mu)}$.
This solution is equivalent to that of Eq.~(\ref{ex_H}) if we associate $\mu=2\beta/(1-\beta)$,
in agreement with~\cite{wm,equi}.
On the other hand, the condition $\Lambda\propto\epsilon_m$ leads in $f(R)$ gravity to
\begin{equation}
\frac{\tilde{R}f'-f}{f'(\tilde{R}f'-2f)}=\mbox{const},
\label{ex_eps}
\end{equation}
which has no power-function solutions and is not equivalent to the law
$\Lambda\propto H^2$.

Finally, we should mention that the decay law~(\ref{genH}) predicted by
$f(R)$ gravity resembles the form used in~\cite{tah},
$\Lambda\propto R=-6H^2(1-q)/c^2$.
This form is also equivalent, in general-relativistic cosmology, to the
condition $\Lambda\propto H^2$~\cite{wm}.
Ref.~\cite{wm} showed that almost all the current phenomenological models of
decaying vacuum (dark energy) can be unified in the form of the modified
matter expansion rate, $\epsilon_m\propto a^{-3+\varepsilon}$,
where $\varepsilon=\mbox{const}$~\footnote{
The condition $\varepsilon=\mbox{const}$ leads in $f(R)$ gravity
to a complicated, generally, differential equation for the function $f(R)$.}.
The relation $\Lambda\propto H^2$ requires $\varepsilon>1$ to get a currently
accelerating universe.
For example, the results of~\cite{int2} give $\varepsilon_0=1.85$~\footnote{
The model with $\gamma\propto r$ used in~\cite{cw,delt} is consistent
with astronomical observations for $|\varepsilon_0|<0.3$.}.
However, this inequality leads also to an accelerating expansion of the
matter dominated universe.
Overall, the type Ia supernovae, cosmic microwave background, and large-scale
structure observations give the constraint $\varepsilon<0.1$, which rules
out the law $\Lambda\propto H^2$~\cite{hs,wm}.

\section{The $R-1/R$ gravity}

The consistence of $f(R)$ models with cosmological data has been studied for
both metric and Palatini variational formalisms~\cite{obs}.
We examine a particular case
\begin{equation}
f(R)=R-\frac{\alpha^2}{3R},
\label{1/R}
\end{equation}
where $\alpha$ is a
constant, which is the simplest way to introduce current cosmic acceleration
in $f(R)$ gravity~\cite{acc}.
This model is referred to as the $R-1/R$ gravity,
and appears (with Palatini variation and in the ECF) to be
compatible with cosmological data, although more observations are
necessary to put stronger constraints on the analyzed parameters~\cite{jerk}.
The current value of $\tilde{R}$ equals $\tilde{R}_0=(-1.05\pm0.01)\alpha$~\cite{Niko2},
which gives the present value of $\gamma$,
\begin{equation}
\gamma_0=0.015\pm0.005.
\label{gamnow}
\end{equation}
Such a small value indicates that the interaction between matter and dark energy
is {\em weak}, as compared to the direct interaction between dark energy and spacetime.
A large error arises from the sensitivity of the term $Rf'(\tilde{R})-2f(\tilde{R})$
around $\tilde{R}_0$.

Similarly, we find the value of $\gamma$ at the deceleration-to-acceleration
transition which occurred at $\tilde{R}_t=-\sqrt{5/3}\alpha$~\cite{Niko2}:
\begin{equation}
\gamma_t=\frac{4}{15}.
\label{gamtr}
\end{equation}
The interaction rate drops significantly ($\sim 20$ times) from the time
of the transition to the present, which does not support the assumption of
the constancy of $\gamma$~\cite{xx,int2}.
As the universe approaches asymptotically a de Sitter expansion,
$\tilde{R}$ tends to $-\alpha$ and the interaction rate $\gamma$ between
matter and dark energy decreases to zero.

In the case of the $R-1/R$ gravity, Eq.~(\ref{w}) for the present time gives
\begin{equation}
w_{\Lambda,0}=-1.10\pm0.02.
\label{w1}
\end{equation}
This result is consistent with the observed $w_{\Lambda,0}=-1.02^{+0.13}_{-0.19}$~\cite{Gold}.
We also find the value of $w_\Lambda$ at the deceleration-to-acceleration transition,
\begin{equation}
w_{\Lambda,t}=-\frac{5}{3}.
\label{w2}
\end{equation}
As the universe approaches a de Sitter phase, $w_\Lambda$ tends to the value $-1$.
Such a value corresponds to dark energy which does not interact with other forms of matter.
Similarly, Eq.~(\ref{x2}) reads
\begin{eqnarray}
& & x_0=0.69^{+0.05}_{-0.04}, \\
\label{x3}
& & x_t=\frac{1}{5}.
\label{x4} 
\end{eqnarray}
The observational constraint on the ratio $x$ in the matter era is
$x<3\times10^{-5}$~\cite{x}.
The deceleration-to-acceleration transition occurred when the
universe was, clearly, dominated by dark energy.
As the universe approaches a de Sitter expansion, $x$ tends to $1$.

From Eq.~(\ref{genH}), we obtain the present value for the
cosmological term $\Lambda$. 
Substituting the observed values $H_0=71\pm4\mbox{km s}^{-1}\mbox{Mpc}^{-1}$~\cite{WMAP}
and $q_0=-0.74\pm0.18$~\cite{Gold}, we find
\begin{equation}
\Lambda_0=(1.46^{+0.41}_{-0.35})\times10^{-52}\mbox{m}^{-2}.
\label{value1}
\end{equation}
If we use the $q_0$ predicted by the $R-1/R$ gravity,
$q_0=-0.67^{+0.06}_{-0.03}$~\cite{Niko2},
we arrive at
\begin{equation}
\Lambda_0=(1.38^{+0.20}_{-0.22})\times10^{-52}\mbox{m}^{-2}.
\label{value2}
\end{equation}
The cosmological constant at the deceleration-to-acceleration
transition is given by $\Lambda_t=H^2_t/c^2$.
Its value depends on the form of the function $f(R)$.
For the case~(\ref{1/R}), we use the expressions for $H_t$,
$\alpha$, and $q_0$ found in~\cite{Niko2} to obtain 
\begin{equation}
\frac{\Lambda_0}{\Lambda_t}=0.96\pm0.08.
\label{value3}
\end{equation}
We see that $\Lambda$ has not changed much since the time when $q$
changed the sign.
A significant decrease of the cosmological term must have happened earlier.

Using $\Omega_{m,0}=0.29^{+0.05}_{-0.03}$~\cite{Gold} then we obtain
\begin{equation}
\frac{r_0}{\gamma_0}=27.2^{+7.9}_{-1.4}.
\label{tau1}
\end{equation}
This value differs significantly from that found in~\cite{int2},
$r_0/\gamma_0=0.54$.
For the deceleration-to-acceleration transition (which is in the range of
redshifts examined by~\cite{Gold}), we have $\Omega_{m,t}=4/5$~\cite{Niko2}.
This gives
\begin{equation}
\frac{r_t}{\gamma_t}=15,
\label{tau2}
\end{equation}
which is again much larger than $\sim 1$.
Therefore, the condition relating the dark energy density to
the Hubble parameter together with the assumption of the constancy of the
interaction rate are incompatible with the $R-1/R$ gravity.
This conclusion agrees with the results of~\cite{wm} which disfavor
the decay law $\Lambda\propto H^2$, and the results of~\cite{Maj}
which support $\dot{r}\neq0$. 

Eq.~(\ref{cons2}) can be written as
\begin{equation}
\dot{\epsilon}_m+nH\epsilon_m=0,
\label{cons3}
\end{equation}
where
\begin{equation}
n=3-\gamma(\Omega_m^{-1}-1).
\label{scal1}
\end{equation}
The $R-1/R$ gravity predicts for the present time
\begin{equation}
n_0=2.96\pm0.01.
\label{scal2}
\end{equation} 
For the moment of the deceleration-to-acceleration transition, the
$R-1/R$ gravity gives 
\begin{equation}
n_t=2.93.
\label{scal3}
\end{equation}
The departure of $n$ from $3$ means that matter alone is
not conserved.
The above values of $n$ are smaller than $3$ (unlike $36/11$ found
in~\cite{KA}), indicating that matter is being produced from dark energy.
It has been shown~\cite{Niko2} that the largest deviation from
the standard nonrelativistic matter scaling occurs around the
deceleration-to-acceleration transition, where $\kappa\epsilon_m\sim\alpha$.
The result~(\ref{scal3}) is a numerical estimation of this deviation.
Since $n$ is very close to $3$ at this transition, and 
the difference $n-3$ is negligible in the early universe~\cite{Niko2},
we may say that $n\sim3$ for the entire matter era.
Therefore, the deviation of the growth of the cosmic scale factor in this era
from the standard law $a(t)\sim t^{2/3}$ is very small which
is consistent with WMAP cosmological data~\cite{WMAP}.
The values $n=2.93$ and $n=2.96$ correspond respectively to $\varepsilon=0.07$
and $\varepsilon=0.04$,
in agreement with the observational constraint $\varepsilon<0.1$~\cite{wm}.
Our results show that the $R-1/R$ gravity in the Palatini
variational formalism is a viable theory of gravitation that explains
current cosmic acceleration.
On the other hand, the compatibility of metric $f(R)$ gravity models
with astronomical observations is still an open problem~\cite{vi}.

Finally, we compare the $R-1/R$ gravity with the general observational
constraint on the matter--dark energy interaction, found in~\cite{hl}.
This constraint is given by
\begin{equation}
r+\frac{\dot{r}}{H(1+r)}<\gamma<-\frac{3w_\Lambda r}{1+r}.
\label{con1}
\end{equation}
However, it is easier to use the conditions from which Eq.~(\ref{con1})
was derived.
The first one, $q_0<0$ (acceleration), is satisfied by the model~(\ref{1/R})
which predicts $q_0=-0.67^{+0.06}_{-0.03}$~\cite{Niko2}.
The second one represents our expectation that the ratio of the matter density
to the dark energy density decreases with the evolution of the universe,
$\dot{r}<0$.
In $f(R)$ gravity, this condition yields 
\begin{equation}
\dot{\tilde{R}}\frac{d}{d\tilde{R}}\biggl[\frac{\tilde{R}f'-3f}{f'(\tilde{R}f'-2f)}\biggr]>0,
\label{con2}
\end{equation}
where $\dot{\tilde{R}}$ is given by~(\ref{phidot}).
For the case~(\ref{1/R}), the quantity $\dot{\tilde{R}}$ is positive in the
range of acceleration $[\tilde{R}_t,\tilde{R}_\infty]=[-\sqrt{5/3}\alpha,-\alpha]$.
Therefore, we obtain
\begin{equation}
\frac{d}{d\tilde{R}}\biggl[\frac{\tilde{R}-\frac{2\alpha^2}{3\tilde{R}}}{(1+\frac{\alpha^2}{3\tilde{R}^2})(\tilde{R}-\frac{\alpha^2}{\tilde{R}})}\biggr]>0,
\label{con3}
\end{equation}
which holds so long as $\tilde{R}<-\alpha/3$.
The last inequality is satisfied for both the matter and dark energy
eras of the universe expansion~\cite{Niko1,Niko2}, and so is the
condition $\dot{r}<0$.

\section{Summary}

$f(R)$ gravity provides a relativistically covariant explanation for
a cosmological constant that varies with time and interacts with matter,
basing on the principle of least action.
We analyzed the interaction between matter and dark energy in
$f(R)$ gravity formulated in the Einstein conformal frame.
We used the Palatini variational principle to obtain the field equations
and apply them to a flat, homogeneous, and isotropic universe filled with dust.
We found how the simplest phenomenological models of a variable
cosmological constant are related to $f(R)$ gravity.
Particularly, $f(R)$ gravity predicts, for a flat universe without
pressure, a simple relation $\Lambda c^2=H^2(1-2q)$.

For the particular case $f(R)=R-\alpha^2/(3R)$,
we found that the interaction rate changes significantly between
the moment of the deceleration-to-acceleration transition and now.
During the same period, the cosmological constant does not
change much, indicating that its significant decrease must have happened
earlier. 
The predicted value of the current interaction rate is on the order of
$1\%$ of the present value of the Hubble parameter, which means that this
interaction is relatively weak.
Consequently, the energy density scaling only slightly (by $\sim1\%$)
deviates from the standard scaling for nonrelativistic matter, which is
consistent with cosmological data.
Therefore, $f(R)$ gravity in the Palatini formalism appears as a viable theory
that explains current cosmic acceleration. 

All our predictions for nondimensional cosmological quantities in the
$f(R)=R-\alpha^2/(3R)$ gravity are independent of the value of
the only parameter in this model, $\alpha$.
This is not true for more complicated cases where a function $f(R)$
contains two or more parameters from which one can compose one or more
nondimensional combinations.

\end{document}